\documentstyle[prl,aps,twocolumn,psfig]{revtex}

\def\Ci{{\rm Ci}}

\begin{document}

\twocolumn[\hsize\textwidth\columnwidth\hsize\csname
@twocolumnfalse\endcsname

\draft

\title{Universal conductance reduction in a quantum wire}
\author{Er'el Granot \footnotemark}

\address{Department of Electrical Engineering, College of Judea and Samaria, Ariel 44837, Israel}

\date{\today}
\maketitle
\begin{abstract}
\begin{quote}
\parbox{16 cm}{\small
Even a single point defect in a quantum wire causes a conductance
reduction. In this paper we prove (without any approximations)
that for {\em any} point impurity this conductance reduction in
{\em all} the sub-bands is \emph{exactly} $2e^2/h$. Moreover, it
is shown that in the case of a surface defect, not only is the
conductance minimum independent of the defect characteristics, but
the transmission matrix also converges to universal
(defect-independent) values. We also discuss particle confinement
between two arbitrarily weak \emph{point} defects.

}
\end{quote}
\end{abstract}

]

\narrowtext \footnotetext{erel@yosh.ac.il} \noindent The
conductance of a quantum wire is quantized. When the quantum wire
is contaminated by impurities, this quantization disappears.
However, at the sub-bands' threshold energies, there is always a
reminiscence of the quantization. Namely, when the Fermi level is
at the band bottom (the threshold energy) of the $(m+1)$th
sub-band, the conductance is exactly $2me^2/h$. This feature is
independent of the impurities' strength. Many works validate this
finding \cite{Chu_Sorbello_89,Bagwell_90,Levinson_etal}. Recently,
however, it has been found
\cite{Chu_Sorbello_89,Bagwell_90,Levinson_etal,Tekmab_Ciraci_90}
 that below the threshold energy of the $m$th mode, even a single
defect is responsible for an unexpectedly large reduction in
conductance: the weaker the impurity, the closer the dip to the
threshold energy. The proximity to the pinned (i.e.,
defect-independent) energy level makes this reduction quite odd.
Chu and Sorbello\cite{Chu_Sorbello_89} attributed this feature to
multiple scattering between the impurity and the wire's
boundaries, and this is discussed in more detail by N\"{o}ckel and
Stone \cite{Nockel_Stone}.

\bigskip

In this paper we calculate the conductance reduction
\emph{exactly} and prove that it is always (for any point defect
and in all the sub-bands) equal to exactly $2e^2/h$. This
reduction is distributed \emph{unevenly} among the propagating
modes.

For such a system (a point impurity in a quasi-1D wire, see
Fig.1), the 2D Schr\"{o}dinger equation is

\begin{equation}
\nabla^{2}\psi+\left( \omega-V
\right)\psi=-D\left(\mathbf{r}-\mathbf{r}_0 \right) \psi
\label{schr_eq}
\end{equation}(hereinafter we use the units $\hbar=2m_0=1$, where $m_0$ is the
electron's mass). $V$ is the potential of the wire walls ($V=0$
inside the wire and $V=\infty$ outside it), $D$ is the defect
potential and $\mathbf{r}_0=\varepsilon \hat{y} $ is the impurity
location. Since the defect has the properties of a point-like
impurity, the right-hand term of the Schr\"{o}dinger equation can
be written $D \left( \mathbf{r}-\mathbf{r}_0 \right)\psi
\left(\mathbf{r}_0 \right)$ \cite{Azbel_91}, which allows for an
exact scattering solution.

Let us denote the incident wave by $\psi_{inc}$. Then, taking
advantage of the point-like nature of the impurity, the scattered
wave function due to the defect's presence can be written
\cite{Granot_Azbel}

\begin{equation}
\psi _{sc}=\psi _{inc}-\frac{\psi _{inc}\left( \mathbf{r}_{0}\right) \int d%
{\mathbf{r}}^{\prime }D\left( \mathbf{r}^{\prime }\mathbf{-r}_{0}\right) }{%
1+\int d{\mathbf{r}}^{\prime }G^{+}\left( \mathbf{r}^{\prime }\mathbf{,r}%
_{0}\right) D\left( \mathbf{r}^{\prime }\mathbf{-r}_{0}\right)
}G^{+}\left( \mathbf{r,r}_{0}\right)
\label{gen_solution}
\end{equation}

where $G^{+}\left( \mathbf{r_1,r_2}\right)$ is the "outgoing" 2D
Green function of the geometry (the wire) and $\psi_{inc}$ is the
incident wave (a homogeneous solution). It should be noted that
eq. \ref{gen_solution} is an exact solution. However, if the
impurity were not an ideal \emph{point} impurity, this equation
would be a first-order approximation in the asymptotic solution
$|\mathbf{r}| \rightarrow \infty$. The Green function for the
given wire geometry takes the form:

\begin{equation}
  G \left( \mathbf{r,r'} \right)=i\sum_{n=1}^{\infty}\frac{\sin(n \pi y) %
  \sin(n \pi y')}{k_n} e^{i k_n |x-x'|}
  \label{green}
\end{equation} where ${\mathbf{r}} \equiv x \hat{x}+y \hat{y}$, ${\mathbf{r'}}
\equiv x' \hat{x}+y' \hat{y}$ and $k_n \equiv
\sqrt{\omega-(n\pi)^2}$ is the effective wavenumber. Hereinafter,
the length parameters are normalized to the wire's width.

\begin{figure}
\psfig{figure=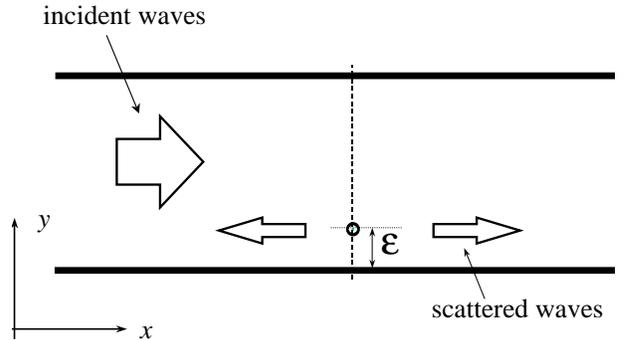,width=8cm,bbllx=160bp,bblly=80bp,bburx=690bp,bbury=470bp,clip=}
\caption{A 2D wire with a single point defect (the black dot)}
\label{fig1}
\end{figure}

\bigskip
Choosing the right potential for the impurity is a very tricky
business as can be understood from the literature
\cite{Chu_Sorbello_89,Levinson_etal,Tekmab_Ciraci_90,Azbel_91,Bagwell_Lake_92,Kim_satanin_99}
(see also ref.17 of \cite{Bagwell_Lake_92}). A simple 2D delta
function (2DDF), which is a natural candidate to represent a point
impurity (like in 1D), i.e., $\delta \left( x\right) \delta \left(
y\right)$, does not scatter (its cross section is zero), and
therefore cannot be used.

Throughout this article we use the Impurity D Function (IDF) that
was first presented by Azbel \cite{Azbel_91}. However, since in
our wire's geometry the problem's symmetry is Cartesian rather
than radial, we choose the following IDF:

\begin{equation}
D\left( \mathbf{r}\right) \equiv  \lim_{\rho \rightarrow 0}\frac{2%
\sqrt{\pi }\delta \left( x\right) }{\rho \ln \left( \rho /\rho _{0}\right) }%
\exp \left( -y^{2}/\rho ^{2}\right) .  \label{point_imp}
\end{equation}

Unlike the 2DDF, this potential, which is infinitely shallower,
does scatter \cite{Azbel_91}. The de Broglie wavelength of the
impurity's bound state is $\lambda_{B}=\pi \rho_0 \exp (\gamma
/2)/2$. This is the only parameter that characterizes the
impurity, and therefore eq. \ref{point_imp} can be used to mimic
\emph{any} impurity with the same de Broglie wavelength, the width
of which is much smaller than $\lambda_{B}$ (i.e., a point
defect).

\bigskip
On the face of it, the solution is straightforward: all that is
needed is to substitute eqs. \ref{point_imp} and \ref{green} into
eq. \ref{gen_solution}. However, the Green function has a
logarithmic singularity at $|\mathbf{r-r'}|\rightarrow 0$. Here is
where the impurity's width $\rho$ plays a major part, and the
limit $\rho \rightarrow 0$ should be approached with great
caution. Therefore, we first solve the integral for a finite
$\rho$ and only then evaluate the limit.

We assume that the incident wave is the $n$th mode, and that the
incident energy is close to the $m$th threshold energy (i.e.,
$\omega \simeq (m \pi)^2$ ), giving

\begin{equation}
\psi_{inc} \left( \mathbf{r} \right)=\sin (n \pi y) \exp \left( i
k_n x \right). \label{incident}
\end{equation}

By using the following relation

\begin{equation}
\begin{array}{r}
  \int dy \sin(n \pi y) \exp \left[
-(y-\varepsilon)^2/\rho^2 \right] = \\
   \rho \sqrt{\pi} \sin(n \pi \varepsilon) \exp \left[-(n \pi \rho/2)^2 \right]
\\
\end{array}
\label{intlimit}
\end{equation}

we find the solution ($x>0$)
\begin{equation}
\psi_{sc} \left( \mathbf{r} \right)=\sum_{l=1}^{\infty} \left(
\delta_{nl}-A_{nl} \right) \sin(l \pi y) \exp \left( i k_l x
\right) \label{Fsolution}
\end{equation}

where $\delta_{nl}$ is the Kronecker delta,

\begin{equation}
A_{nl} \equiv \frac{\sin (n \pi \varepsilon) \sin (l \pi
\varepsilon)}{i k_l \left[ \frac{\ln(\rho_0/\bar{\rho})}{2
\pi}+\sum_{n'\leq m} \frac{\sin^2(n' \pi \varepsilon)}{i k_{n'}}
\right]} \label{A_def}
\end{equation}

and $\bar{\rho}$ is some length scale which depends on the
impurity's location ($\varepsilon$), the incident energy $\omega$
and $m$:

\begin{equation}
\ln ( \bar{\rho} ) \equiv \lim_{\rho \rightarrow 0} \left\{ \ln
\rho+2 \pi \sum_{n'=m+1}^{\infty} \frac{\sin^2 (n' \pi
\varepsilon)}{q_{n'}} e^{-(n' \pi \rho/2)^2 } \right\}
\label{rho_bar}
\end{equation}

where $q_n \equiv \sqrt{(n\pi)^2-\omega}$.

The conductance can be evaluated by the Landauer equation
\cite{Landauer}

\begin{equation}
G=\frac{1}{\pi}\sum_{n,l<m}T_{nl} \label{Landauer}
\end{equation}

where

\begin{equation}\label{trans_dia}
  T_{nl}=\left\{
  \begin{array}{ll}
    |1-A_{nn}|^2 & n=l \\
    |A_{nl}|^2\frac{k_n}{k_l} & n \neq l \
  \end{array}
  \right.
\end{equation}

are the transmission coefficients. Clearly, at the threshold
energies where $\omega=(m\pi)^2$, the coefficients vanish,
$A_{nl}=0$ for any $n,l<m$, and therefore

\begin{equation} \label{thre_cond}
G=\frac{1}{\pi}(m-1)
\end{equation}

independent of the impurity, as has been shown in previous works
\cite{Chu_Sorbello_89,Bagwell_90,Levinson_etal}.

However, eq. \ref{A_def} allows us to calculate the \emph{minima}
of the conductance as well. The minima are obtained when the
imaginary part of eq. \ref{A_def} vanishes, i.e., when

\begin{equation} \label{minim_re}
\frac{\ln(\rho_0/\bar{\rho})}{2\pi}=\frac{\sin^2(m\pi\varepsilon)}{q_m},
\end{equation}

and thus

\begin{equation}
A_{nl}^{min} = \left[ \sum_{j<m}\frac{\sin^2(j \pi
\varepsilon)}{\sin(n \pi \varepsilon) \sin(l \pi
\varepsilon)}\frac{k_l}{k_j} \right]^{-1}. \label{Anl_min}
\end{equation}

Using eqs. \ref{Landauer}, \ref{trans_dia} and \ref{Anl_min}, at
the minimum points ($G=G_m$)

\begin{equation}
\begin{array}{l}
G_{m}=\\ \frac{1}{\pi}\sum_{n,l<m} \left\{ \delta_{nl}- \left[
\sum_{j<m}
\frac{\sin^2(j\pi\varepsilon)}{\sin(n\pi\varepsilon)\sin(l\pi\varepsilon)}
\frac{k_l}{k_j} \right]
\right\}^2\\
\end{array}
\label{cond_all}
\end{equation}

This complicated expression can be considerably simplified: with
the following definition

\begin{equation}
\label{sigma} \sigma \equiv \sum_{j<m}\frac{\sin^2(j\pi
\varepsilon)}{k_j}
\end{equation}

eq. \ref{cond_all} can be rewritten

\begin{equation} \label{cond_with_s}
\begin{array}{r}
\pi G_{m}=\sum_{n<m}1-\frac{2}{\sigma} \sum_{n<m}
\frac{\sin^2(n\pi\varepsilon)}{k_n}+\\
\frac{1}{\sigma^2}\sum_{n,l<m}\frac{\sin(l\pi\varepsilon)\sin(n\pi\varepsilon)}{k_lk_n}.\\
\end{array}
\end{equation}

The first term of eq. \ref{cond_with_s} is equal to $m-1$, the
second and third terms are, by definition, equal to $2$ and $1$,
respectively, and therefore the minimum conductance near the $m$th
threshold energy is simply

\begin{equation} \label{con_min_final}
G_m=\frac{1}{\pi} (m-2)
\end{equation}

which, again, is independent of the defect's properties (location
and strength). \emph{Any} point defect will exhibit the same
conduction transition, from the minima (eq. \ref{con_min_final})
to the maxima (eq. \ref{thre_cond}). Hence, the defect reduced the
conductance by \emph{exactly} $\Delta G\equiv G_{max}-G_{min}=
\pi^{-1}$ ($=2e^2/h$) in every band (see Fig. 2).

\begin{figure}
\psfig{figure=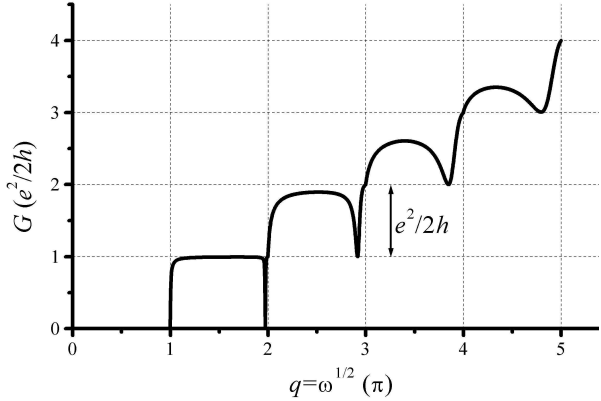,width=8cm,bbllx=40bp,bblly=10bp,bburx=530bp,bbury=380bp,clip=}
\caption{A plot of the conductance as a function of the normalized
Fermi wavenumber (units of $\pi$)} \label{fig2}
\end{figure}

In general, eq. \ref{A_def} is considerably simplified when the
defect is close to the boundary, i.e., $\varepsilon \ll 1$. In
that case (near the $m$th band)

\begin{equation} \label{A_for_small_ep}
A_{nl} \simeq \left[ \sum_{j<m}{ \frac{j^2}{nl}
\sqrt{\frac{m^2-l^2}{m^2-j^2}} + i \pi \frac{\sqrt{m^2-l^2}}{nl}
\left(\Delta^{-1}-\frac{m^2}{q_m}  \right)} \right]^{-1}
\end{equation}


and

\begin{equation} \label{two_def}
\Delta \equiv 2 \pi \frac{(\pi \varepsilon)^2}{ln(\rho_0 /
\varepsilon C )}
\end{equation}

where $C \equiv 4\exp[\gamma/2-\Ci(\pi)] \simeq 5$ is a numerical
constant ($\gamma$ is the Euler constant and $\Ci$ is the cosine
integral).

Hence, the reduction takes place at the following energies

\begin{equation} \label{red_energy}
\omega_m^*\simeq (m \pi)^2-m^4 \Delta^2
\end{equation}

These energies depend on the defect's characteristics (via
$\Delta$) but the \emph{amount of the reduction} does not.

In the limit $\varepsilon \rightarrow 0$, i.e., when the defect is
a surface defect, another universality appears: not only is the
conductance independent of the defect characteristics but the
transmission matrix is also defect-independent. At the minima, the
transmission coefficients converge to the limits

\begin{equation} \label{T1_limit}
\lim_{\varepsilon \rightarrow 0} T_{nn}=\left| 1- \left[
\sum_{j<m} \frac{j^2}{n^2} \sqrt{\frac{m^2-n^2}{m^2-j^2}}
\right]^{-1} \right|^2
\end{equation}

\begin{equation} \label{T2_limit}
\lim_{\varepsilon \rightarrow 0} T_{l \neq m}= \left[ \sum_{j<m}
\frac{j^2}{nl} \sqrt{\frac{m^2-l^2}{m^2-j^2}} \right]^{-2}
\sqrt{\frac{m^2-n^2}{m^2-l^2}}
\end{equation}

which are merely pure numbers, in which no reminiscence of the
defect's characteristics is left out.

The fact that the reduction in conductance is independent of the
band number is not trivial since this reduction is distributed
\emph{unevenly} among all the propagating modes. Therefore, the
first reduction, just below the $m=2$ sub-band is a special case.
At this energy the conductance is reduced to zero, and is a
consequence of a \emph{single} mode (the first one), which is
totally reflected. This is the only point at which a propagating
mode is totally reflected by a \emph{single} point defect (note
that unlike ref. \cite{Gurvitz_Levinson}, we obtained this result
without any approximations). This unique energetic place can be
used to create a bound state in the continuum, simply by binding
it between two totally reflecting defects. At this energy Kim and
Satanin \cite{Kim_satanin_99} also found a bound state in the
continuum, but for the problematic 2DDF model. The presence of an
additional defect can create zero transmission regions at higher
energies, and therefore with more defects it becomes possible to
bind particles at higher sub-bands. In the following we will
evaluate the minimum distance between two defects, which allows
for such binding.

\begin{figure}
\psfig{figure=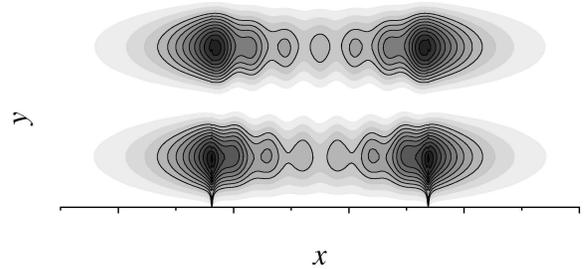,width=8cm,bbllx=40bp,bblly=90bp,bburx=500bp,bbury=320bp,clip=}
\caption{The probability density of the bound state} \label{fig3}
\end{figure}

By adding another defect, the Schr\"{o}dinger equation will look
like

\begin{equation}
\nabla^{2}\psi+\left( \omega-V
\right)\psi=-D\left(\mathbf{r}-\mathbf{r}_1 \right)
\psi-D\left({\mathbf{r}}-{\mathbf{r}}_1-L\hat{x} \right) \psi.
\label{schr_eq_two_defects}
\end{equation}

It is then clear that if the defects are located at the same
distance from the boundary and the distance between them, $L$, is
extremely large ($L \gg 1$ so the inter-scattering can be ignored)
and maintains $L=\frac{n \pi}{\sqrt{\omega_2^*-\pi^2}} \simeq
\frac{n}{\sqrt{3}}$ (where $n$ is an integer, and $\omega_2^*$ can
be evaluated from eq. \ref{red_energy} for $m=2$ when the defects
are very far apart $n \rightarrow \infty$), the system will hold a
bound eigen state with the eigen energy that corresponds to
$\omega_2^*$ (when the distance between them is finite, the
binding energy increases). In Fig. 3, the probability density of
such a bound eigenstate (for $\varepsilon=10^{-3}$) is shown.

It is well known that in wave dynamics only an infinite barrier
(either high or long) can totally reflect the incident wave.
Therefore, in principle, only infinite barriers can confine a
quantum state. So how is it possible for only two point
scatterers, which can be arbitrarily weak, to confine an
energetically bound state in the continuum? The answer is that
what really confines the quantum particle is the infinite wire's
boundaries rather than the point scatterers. All the scatterers
need to do is to deviate the particle's trajectory a little .

\begin{figure}
\psfig{figure=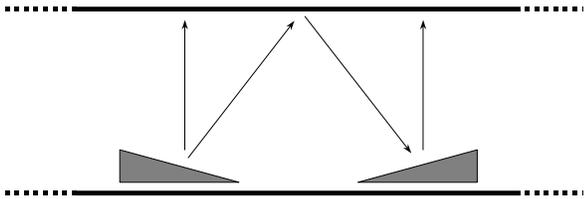,width=8cm,bbllx=150bp,bblly=200bp,bburx=650bp,bbury=500bp,clip=}
\caption{Confinement of a \emph{classical} particle between two
opposite slopes. The point defects in the quantum case behave like
these slopes} \label{fig4}
\end{figure}

A classical analogy to such a confinement is shown in Fig. 4. Two
opposite planes, which have an arbitrary small slope, can confine
an energetic classical particle between them by directing the
particle directly towards the wire's boundaries.

One should therefore expect that the strength of the defect will
determine a minimum distance within which it is possible to
confine a particle.

After some tedious, albeit straightforward calculations, it can be
shown that a bound state in the $M$th propagating sub-band (i.e.,
$[(M+1)\pi]^2>\omega>(M\pi)^2$) should satisfy the following
coupled equations (the second equation holds only in the limit
$\rho \rightarrow 0$):

\begin{equation}
\sum_{n=1}^{M}\frac{\sin^2(n\pi\varepsilon)}{k_n}\sin^2(k_nL/2)=0
\label{req1}
\end{equation}

\begin{equation}
\frac{\ln(\rho_0/\rho)}{2\pi}=\sum_{n=M+1}^{\infty}\frac{\sin^2(n\pi\varepsilon)}{q_n}\left[
e^{-(n\pi\rho/2)^2}-e^{-q_nL}\right]. \label{req2}
\end{equation}

Since all the terms in eq. \ref{req1} are non-negative, a solution
is possible only when all the terms vanish. For an arbitrary
$\varepsilon$ (non-rational) we should demand
\begin{equation}
\begin{array}{c}
  \sqrt{\omega-\pi^2}L=2n_1\pi, \\
  \sqrt{\omega-(2\pi)^2}L=2n_2\pi, \\
  \vdots \\
  \sqrt{\omega-(M\pi)^2}L=2n_M\pi, \\
\end{array}
\end{equation}

where $n_1,n_2,\ldots n_M$ are integers. Eq. \ref{req2} is
reduced, in the limit of a surface defect $\varepsilon \rightarrow
0$, to a simple expression

\begin{equation}
L=-\frac{\ln \left\{1-q_{M+1}/(M+1)^2\Delta \right\}}{q_{M+1}}.
\end{equation}

Hence, the minimum distance within which it is possible to confine
the bound state is inversely proportional to $\Delta$:
\begin{equation}
L_{min}=1/(M+1)^2\Delta.
\end{equation}

To summarize, we calculated the transmission and conductance of a
quantum wire which was contaminated by a single defect. We showed
that the reduction in the conductance in all the sub-bands is
\emph{totally universal} (independent of the defect's
characteristics), and is \emph{always} equal to $2e^2/h$.
Moreover, we showed that when the point defect is a surface
impurity, the transmission coefficients at the minima converge to
\emph{ universal numbers} (and again, are independent of the
defect). We used this result to show that it is possible to
confine an eigenstate in the \emph{continuum} of the quantum wire
between two totally reflecting defects, and to show the
limitations enforced upon it.

It should be stressed that while the discussion was focused on
quantum wires, this effect can occur in any waveguide with a
single point scatterer: acoustical waveguide, electromagnetic
waveguide, optical waveguide, etc.

\end{document}